\documentclass{ieeeaccess}
\usepackage{cite}
\usepackage{amsmath,amssymb,amsfonts}
\usepackage{algorithmic}
\usepackage{graphicx}
\usepackage{textcomp}
\usepackage{siunitx}
\usepackage{subfig}
\usepackage{csvsimple}
\usepackage{xcolor}
\usepackage{makecell}

\usepackage{bm}
\makeatletter
\AtBeginDocument{\DeclareMathVersion{bold}
\SetSymbolFont{operators}{bold}{T1}{times}{b}{n}
\SetSymbolFont{NewLetters}{bold}{T1}{times}{b}{it}
\SetMathAlphabet{\mathrm}{bold}{T1}{times}{b}{n}
\SetMathAlphabet{\mathit}{bold}{T1}{times}{b}{it}
\SetMathAlphabet{\mathbf}{bold}{T1}{times}{b}{n}
\SetMathAlphabet{\mathtt}{bold}{OT1}{pcr}{b}{n}
\SetSymbolFont{symbols}{bold}{OMS}{cmsy}{b}{n}
\renewcommand\boldmath{\@nomath\boldmath\mathversion{bold}}}
\makeatother

\def\BibTeX{{\rm B\kern-.05em{\sc i\kern-.025em b}\kern-.08em
    T\kern-.1667em\lower.7ex\hbox{E}\kern-.125emX}}

\begin{document}
\history{Date of publication xxxx 00, 0000, date of current version xxxx 00, 0000.}
\doi{10.1109/ACCESS.2023.1120000}

\title{Imageless Contraband Detection Using a Millimeter-Wave Dynamic Antenna Array via Spatial Fourier Domain Sampling}

\author{Daniel Chen, \IEEEmembership{Graduate Student Member, IEEE},
Anton Schlegel, \IEEEmembership{Graduate Student Member, IEEE}, and Jeffrey A. Nanzer,
\IEEEmembership{Senior Member,~IEEE}}

\address{Department of Electrical and Computer Engineering,
	Michigan State University, East Lansing, MI 48824 USA (email:
	chendan7@msu.edu, schleg19@msu.edu, nanzer@msu.edu)}

\tfootnote{Manuscript received April 2024. This work was supported in part by the National Science
	Foundation under Grant 1751655.}

\markboth
{Chen \headeretal: Imageless Contraband Detection Using a Millimeter-Wave Dynamic Antenna Array via Spatial Fourier Domain Sampling}
{Chen \headeretal: Imageless Contraband Detection Using a Millimeter-Wave Dynamic Antenna Array via Spatial Fourier Domain Sampling}

\corresp{Corresponding author: Jeffrey A. Nanzer (e-mail: nanzer@msu.edu).}

\begin{abstract}
	We demonstrate an imageless method of concealed contraband detection using a real-time 75~GHz rotationally dynamic antenna array. The array measures information in the two-dimensional Fourier domain and captures a set of samples that is sufficient for detecting concealed objects yet insufficient for generating full image, thereby preserving the privacy of screened subjects. The small set of Fourier samples contains sharp spatial frequency features in the Fourier domain which correspond to sharp edges of man-made objects such as handguns. We evaluate a set of classification methods: threshold-based, K-nearest neighbor, and support vector machine using radial basis function; all operating on arithmetic features directly extracted from the sampled Fourier-domain responses measured by a dynamically rotating millimeter-wave active interferometer. Noise transmitters are used to produce thermal-like radiation from scenes, enabling direct Fourier-domain sampling, while the rotational dynamics circularly sample the two-dimensional Fourier domain, capturing the sharp-edge induced responses. We experimentally demonstrate the detection of concealed metallic gun-shape object beneath clothing on a real person in a laboratory environment and achieved an accuracy and F1-score both at 0.986. The presented technique not only prevents image formation due to efficient Fourier-domain space sub-sampling but also requires only 211~ms from measurement to decision.
\end{abstract}


\titlepgskip=-21pt

\maketitle
%%%%%%%%%%%%%%%%%%%%%%%%%%%%%%%%%%%%%%%%%%%%%%%%%%%%%%%%%%%%%%%%%%%%%%%%%%%%%%%%%%%%%%%%%%%%%%%%%%%%%%%%%%%%%%%%%%%%%%%%%
\section{Introduction}
\PARstart{R}{emote} detection of objects is an important function for many millimeter-wave applications, including concealed contraband detection~\cite{4665741,bhu2du}, human-machine interfacing~\cite{lien2016soli,hazra2018robust}, airborne remote sensing~\cite{4721105, 6809937}, and environmental sensing for autonomous vehicles~\cite{rasshofer2005automotive,hasch2012millimeter}, among many others. The characteristics of millimeter-wave signals are desirable for imaging and sensing applications since the wavelengths are sufficiently short to achieve adequate image resolution while also long enough to propagate through fog, smoke, garments, and other obscurants with negligible attenuation~\cite{Currie1987,Nanzer2012}. In applications of concealed contraband detection where the primary background is a person, the ability to identify potential prohibited objects without compromising personal privacy can be beneficial. \textcolor{black}{With recent advancements of automatic recognition techniques,  it is possible to extract sensitive information from a wide range of biometric modalities~\cite{9481149}; for example, an individual's age and gender can be determined from gait patterns~\cite{9423216,9870842,10386667}. Imagery from security screenings of people falls in the category of biometric data that potentially contains personal information that may be used for malicious purposes.} Typically, image-based contraband detection relies on first forming full images of the measured scene after which image-based signal processing techniques are applied for detection and classification~\cite{4665741,bhu2du}. This approach, while functional, has two potential areas for improvement: 
\begin{enumerate}
	\item Privacy, where sensitive information of the screened person are revealed; and
	\item Efficiency, which considers whether full images are required, since typical prohibited objects rarely occupy a significant fraction of the recovered images.
\end{enumerate}

Approaches addressing the above two points are therefore of interest. \textcolor{black}{Recent research has sought to address the challenges related to privacy issues with such sensing techniques. As discussed in~\cite{9481149}, there are multiple points of \textcolor{black}{opportunity} where privacy enhancing techniques can be applied, ranging from designing imaging sensors with embedded privacy protection features~\cite{4270411}, using template protection techniques~\cite{7192825}, \textcolor{black}{de-identifying} sensitive information within the data~\cite{1377174, 7285017}, using cancelable biometrics~\cite{7192838}, sharing data using privacy-preserving schemes~\cite{1377174}, and applying adversarial approaches for automatic recognition techniques~\cite{8744230}, among others. In general, these approaches can be categorized into three levels~\cite{9481149}: image level, representation level, and inference level. Regardless of where a privacy enhancing technique is applied, there generally exists a trade space between privacy enhancement and biometric utility of the measured data. This means that complete privacy protection can eliminate the utility of biometric data while a complete biometric utility offers no protection on privacy. One possible solution to this problem is to shift away from using modalities that require imaged-based data.} \textcolor{black}{It has been successfully demonstrated that imageless remote sensing systems are adequate to differentiate the presence or absence of concealments that can either be made from metallic and/or dielectric materials~\cite{6132255, Kapilevich_Harmer_Bowring_2017, 10501605}. One possible challenge is, however,} without recovering images, the lack of sufficient spatial information can necessitate alternative processing techniques whereby the sensing and/or detection functions take place in a different domain or dimension. One promising approach, demonstrated in~\cite{9660363,9704178,9865391}, is to directly measure the scene in the spatial frequency domain, i.e., the Fourier transform of the spatial domain. The spatial frequency domain can contain broad spatial-spectral responses generated by sharp edges of objects that manifest at spatial frequency domain angles normal to the edges in the spatial domain. In particular, the spatial frequency responses manifesting from common man-made structures can be captured by only sampling a small fraction of the Fourier-domain information, substantially less than the required amount of data for full image reconstruction~\cite{9761202}. In~\cite{9395397}, a ring-shaped spatial frequency filter was generated by a two-element interferometric antenna array with a rotational trajectory relative to the array centroid, from which the classification of ground scene images with and without man-made structures was demonstrated. One shortcoming with this approach is the measured scenes' relative orientation to the dynamic antenna array which affects back-scattered signals incident on the receivers. This may yield wide variations in the measured Fourier-domain responses, adversely impacting classification performance.

\begin{figure*}[t!]
	\centering
	\includegraphics[width=\textwidth]{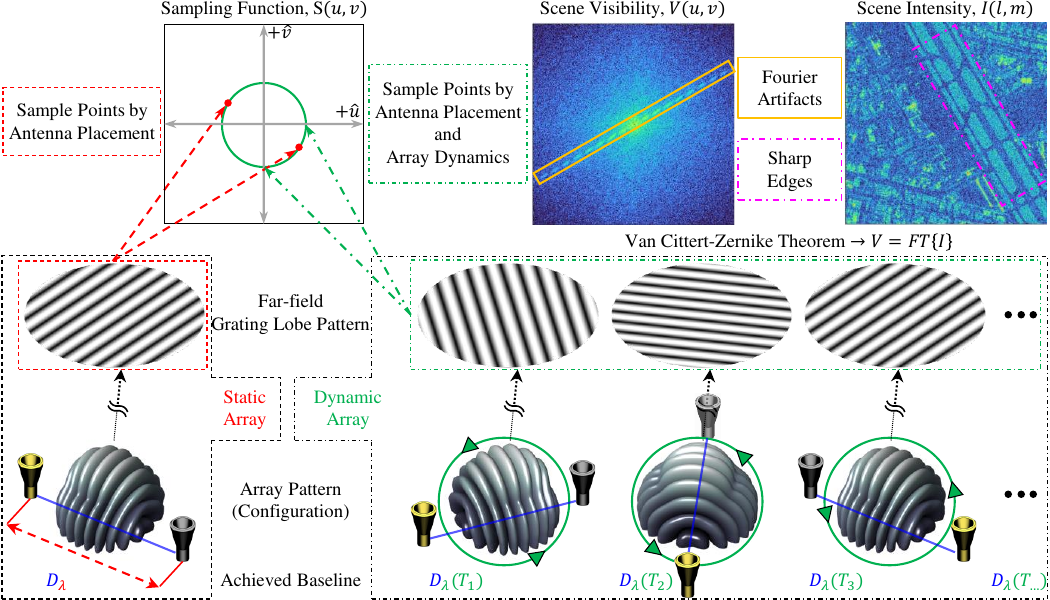}
	\caption{Overview of the interferometric technique and the comparison between static and dynamic antenna array. The smallest element of an interferometric array is a correlation pair which is shown by the yellow and gray conical horn antennas. Conventional interferometric arrays are considered static where the array pattern of a given correlation pair within the array is defined by antenna placement achieving a baseline, $D_\lambda$, and a far-field grating lobe pattern that corresponds to particular sampling points in the Fourier domain. The sampling Fourier-domain space is also known as the scene visibility, $V(u,v)$, which is the two-dimensional Fourier transformation of the scene intensity, $I(l, m)$, when satisfying the Van Cittert-Zernike theorem~\cite{Born:1999ory,Thompson2017} where the fields radiated from the measured scene \textcolor{black}{are} considered \textcolor{black}{spatio-temporally} incoherent. The visibility is the collection of all spatial frequency information of the measured scene which are related to particular spatial/physical features. As shown in the upper right corner, boxed by solid orange and dash-dot magenta, sharp edges in the spatial domain manifest radially extending Fourier artifacts that are orthogonal to the edge direction. Such Fourier artifacts can be easily sampled by utilizing array dynamics, assuming the measured scene is relatively static, where the achieved baseline $D_\lambda(t)$ depends on both the antenna placement and the array trajectory. The dynamic antenna array shown above is a correlation pair rotating with respect to their centroid, synthesizing a ring-shaped sampling function over a rotational trajectory of 180$^\circ$ that can effectively capture the radially extending Fourier artifacts in the $uv$-plane. Data source for Scene Intensity~\cite{ICEYE}.}
	\label{fig:OV1}
\end{figure*}

In this work, a \SI{75}{GHz} rotational dynamic antenna array is designed to enable imageless detection of metallic gun-shape object that is concealed beneath clothing on a real person. In contrast to other works, the presented system does not form images, and does not collect sufficient information to form an image. The approach depends on appropriate sampling of the spatial frequency information exhibited by the measured scene, which requires signals radiated and/or scattered from the scene to be spatially and temporally incoherent. To satisfy such signal properties, the rotational dynamic antenna array is implemented in conjunction with the recently demonstrated active incoherent millimeter-wave (AIM) technique~\cite{8458190,8654605,9079644}. In this technique, noise transmitters are used to illuminate the measured scene achieving the aforementioned spatio-temporally incoherent radiating condition while also obtaining a high received signal-to-noise ratio (SNR).

We previously demonstrated the potential for feature separability using a simple threshold detection separating a mannequin concealing metallic gun-shape object under clothing versus a mannequin without contraband with the assumption that the measured scenes were relatively static to the array dynamics~\cite{10111012}. However, this assumption is unlikely to be satisfied regardless of how fast the implemented array dynamics \textcolor{black}{are} when the measured scene involves a real person, which is a more practical scenario involved in security screening application as breathing and/or torso movements can happen at any instance even during the fast array dynamics screening process. In this work we investigate the validity of the imageless contraband detection approach where the involved screening subject is a real person and the considered contraband gun-shape object is metallic by exploring various algorithms applied to the extracted arithmetic features. We further expand on our prior work by discussing the dynamic antenna array theory, presenting a detailed description of the system, and experimentally demonstrating the classification of a person concealing metallic gun-shape object under clothing versus a person without contraband by achieving an accuracy and F1-score both at 0.986. The demonstrated technique required a total time of \SI{221}{ms} from measurements to detection decision which is applicable to real-time contraband detection scenario while preventing image formation of the screened subjects due to the efficient Fourier-domain sub-sampling enabled by the rotational dynamic antenna array.
%%%%%%%%%%%%%%%%%%%%%%%%%%%%%%%%%%%%%%%%%%%%%%%%%%%%%%%%%%%%%%%%%%%%%%%%%%%%%%%%%%%%%%%%%%%%%%%%%%%%%%%%%%%%%%%%%%%%%%%%%
\section{Interferometric Imaging Technique and Sub-sampling Through Array Dynamics}
The interferometric imaging technique is one of various methods that can be implemented at millimeter-wave bands to leverage the wavelength characteristics as well as enabling a relatively compact design compared to lower frequency systems. Interferometric imaging is a Fourier domain approach where sampling of the measured scene information takes place in the spatial frequency domain and the recovery of imagery relies on applying an inverse Fourier operation when sufficient spatial frequency samples are acquired~\cite{Thompson2017,Nanzer2012}. The technique has seen growing interest as it is capable of generating high-resolution imagery while requiring only a fraction of the aperture area necessary in imaging systems based on mechanically- and/or electronically-steered antennas or focal plane arrays~\cite{898661,7446315,8416691}. Spatial frequency samples of a given scene are obtained by cross-correlating the received signals between antenna pairs in an array. The smallest element of an interferometric array is a two-element correlation pair as shown in lower left of Fig.~\ref{fig:OV1}. In general, for a relatively small number of physical elements, the number of spatial frequency samples that can be captured can be large, which is the direct result \textcolor{black}{of} the number of unique antenna pairs in the array. By judicious placement of a small number of antennas, a dense sampling function can be synthesized enabling high-resolution image reconstruction using sparse antenna arrays~\cite{6048955,Thompson2017,6688267,7870618}.

The Fourier space where interferometric imaging systems collect spatial frequency information is known as the scene visibility, $V(u,v)$ (see Fig.~\ref{fig:OV1}), with the variable $u$ and $v$ each representing a dimension of the two-dimensional visibility plane. Furthermore, when the Van Cittert-Zernike theorem~\cite{Born:1999ory,Thompson2017} is satisfied, where the fields radiated from the measured scene are spatially and temporally incoherent, the scene visibility $V(u,v)$ can then be given by the two-dimensional Fourier transformation of the scene intensity, $I(l,m)$ \textcolor{black}{which is a real value function}, with $l=\sin\theta\cos\phi$ and $m=\sin\theta\sin\phi$ representing the direction cosines in the azimuth and elevation plane, respectively. The Fourier relationship between the scene visibility and intensity is thus given by
\begin{equation}\label{eq.vis}
	V(u,v) = \iint\limits_{-\infty}^{+\infty}I(l,m)e^{j2\pi(ul+v\beta)}dl dm\mathrm{.}
\end{equation}
The sampling function $S(u,v)$ of a static interferometric imaging system is synthesized by the array configuration (i.e., number and location of the antennas) where each sample corresponds to particular visibility points as shown in upper left of Fig.~\ref{fig:OV1}. The sampled visibility (the collection of all visibility information due to the array design) can be considered as the product \textcolor{black}{of} the scene visibility and the synthesized sampling function where each sample is the result of cross-correlating the received signals at the corresponding antenna pair in the array. Assuming that an ideal sampling function exists where $S(u,v)=1$, the reconstructed scene intensity $I_{\mathrm{r}}(l,m)$ can then be recovered by applying the inverse Fourier transformation on the sampled visibility $V_s(u,v)=V(u,v)\cdot S(u,v)$,
\begin{equation}\label{eq.ir}
	I_{\mathrm{r}}(l,m) = \iint\limits_{-\infty}^{+\infty}V_s(u,v)e^{-j2\pi(ul+vm)}dudv\mathrm{.}
\end{equation}
Together, (\ref{eq.vis}) and~(\ref{eq.ir}) provide an overview of measuring information in the spatial frequency domain and its relationship to the spatial domain where the desired image information resides. 

\begin{figure}[t!]
	\centering
	\includegraphics[width=\linewidth]{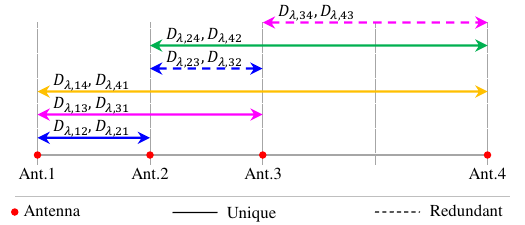}
	\caption{Example of a one-dimensional linear array consists of four antennas demonstrating the concept of unique and redundant sampling point due to the uniqueness of the baseline between two antennas that is determined by their electrical separation and orientation. Each color specifies the unique baseline associated for a given pair of antennas. Based on \textcolor{black}{increasing} antenna numbering label, solid lines represent the considered unique baselines and the dash lines represent the redundant baselines.}
	\label{fig:1d_la_example}
\end{figure}

Practical implementations of interferometric arrays have antennas located at discrete positions, meaning that the sampling function will be discrete, and so will the sampled visibility. Let $x$ and $y$ denote the spatial locations of the receiving antennas, for a two-dimensional spatial frequency plane, the exact $u$ and $v$ locations being sampled due to the pairs of antennas are related through $u=\tfrac{D_x}{\lambda}$~rad$^{-1}$ and $v=\tfrac{D_y}{\lambda}$~rad$^{-1}$, where $D_x$ and $D_y$ represent the antenna separation in the spatial $x$ and $y$ dimensions of a given antenna pair, \textcolor{black}{and} the wavelength of the received radiation is $\lambda=\tfrac{c}{f}$. For a given two-dimensional interferometric array, the complete collection of all discretely sampled $uv$-points (i.e., spatial frequency information due to the array's configuration) can be formulated as
\begin{equation}\label{eq.sf}
	S(u,v) = \sum_{n=1}^{N}\sum_{m=1}^{M}\delta(u-u_n)\delta(v-v_m),
\end{equation}
where $\delta(\cdot)$ is the Dirac delta function and the product $NM$ represents the total possible number of spatial frequency samples that the imaging array can acquire. Note that $NM$ can be considered as the collection of redundant and unique spatial frequency samples
\begin{equation}\label{eq.nm}
	NM = (NM)_{\mathrm{redundant}} + (NM)_{\mathrm{unique}}\mathrm{.}
\end{equation}
\textcolor{black}{
	This is demonstrated in Fig.~\ref{fig:1d_la_example} using a one-dimensional linear array with four antennas. Each color specifies the unique baseline associated with a given pair of antennas. Solid lines represent unique baselines and the dashed lines represent redundant baselines. For example, the pair of Antennas~1 and 2 and the pair Antennas~2 and 3 both \textcolor{black}{yield} the same electrical baseline and orientation (i.e., in the one-dimensional consideration), and hence will sample the same spatial Fourier sampling points. When multiple pairs of antennas within the array synthesize the exact sampled $uv$-points, a redundant sample is obtained and contributes to the term $(NM)_{\mathrm{redundant}}$. Depending on the application, redundant samples may be useful, for example in calibration, but the typical design objective is to optimize $(NM)_{\mathrm{unique}}$ using a fixed number of receiving elements. In general, when the number of receiving elements is small, the receiver numbers tends to have higher positive correlation to $(NM)_{\mathrm{unique}}$. However, challenges arise when trying to implement additional elements, especially for arrays that are large, such as avoiding redundant antenna baselines.}

The interferometric imaging technique requires the signals radiated from the measured scene to satisfy the Van Cittert-Zernike theorem such that they are spatially and temporally uncorrelated~\cite{Born:1999ory,Thompson2017}. This condition can be satisfied either based on the scene's intrinsic thermal radiation~\cite{4381297}, or via active illumination using incoherent signals~\cite{8458190,8654605,9079644}; either will produce incoherence in both space and time to support the Fourier-domain sampling of visibility. Passive interferometric imaging systems usually necessitate implementation with very high sensitivity receivers to compensate the exceedingly small radiated thermal power at millimeter-wave frequencies~\cite{7685,1025064,1388047,4393554}. On the contrary, active systems can leverage transmission of spatio-temporally incoherent signals to illuminate the measured scene, thus increasing the received power. This alleviates requirements for high gain receivers.

Prior works have demonstrated that objects exhibiting discrete spatial responses (i.e., edges) will induce corresponding responses in the visibility space that are broad spectrum, but are located along angles orthogonal to the edges~\cite{9660363,9704178,9865391}. This is also demonstrated in the upper right corner of Fig.~\ref{fig:OV1} by the solid orange and dash-dot magenta boxes that represent the radially extending Fourier artifacts orthogonal to the sharp edges' direction. A Fourier domain ring-shaped filter was introduced in~\cite{9395397} to capture these edge related Fourier artifacts. \textcolor{black}{The ring-shaped filter (i.e., sampling function in the spatial Fourier domain) exhibits a circumference of a circle with its radius commensurate to the receiver baseline, and is being referred to as the ring filter.} The ring filter can be obtained using a simple interferometric array with rotational dynamics based on the assumption that the measured scene is relatively static over the period that the array requires to synthesize the desired sampling function (i.e., ring filter). By introducing array dynamics into the design, the challenge for a static array to improve the quantity $(NM)_{\mathrm{unique}}$ from~(\ref{eq.nm}) can be significantly reduced without requiring additional elements. This is shown in Fig.~\ref{fig:OV1} near the lower right corner where the same correlation antenna pair (as in the static array) is rotated with respect to its centroid tracing out a circular trajectory with three examples shown at various times throughout the array dynamics. Assuming 
uniformly measured intervals at a constant rotational speed, a finer interval results in a less discrete spatial frequency domain ring filter. As expected, when the measured scene does not vary over the duration of a full sampling function generation through array dynamics, additional unique spatial frequency sampling points can be obtained without the need for considering additional receiving elements. 

For a single ring filter, considering rotational array dynamics of duration $T$ of a correlation antenna pair rotating with respect to its centroid, the spatial frequency sampling function (\ref{eq.sf}) becomes
\begin{equation}\label{eq.sf_ring_t}
	S(u,v)= \sum_{t=0}^{T}\delta(u-u(t))\delta(v-v(t))\mathrm{,}
\end{equation}
where 
\begin{equation}\label{eq.u_t}
	u(t) = D_{\lambda}\sin(\gamma(t))\mathrm{,}
\end{equation}
and
\begin{equation}\label{eq.v_t}
	v(t) = D_{\lambda}\cos(\gamma(t))\mathrm{.}
\end{equation}
The quantity $D_{\lambda}$ in~(\ref{eq.u_t}) and~(\ref{eq.v_t}) is the baseline separation of the two antennas relative to the wavelength; $\gamma(t) = \gamma_0 + \gamma_r\tau$ is the angle between the linear pair and reference (e.g., $\gamma_0 = 0^\circ$ for the horizontal plane) as function of time where $\gamma_r$ is the rotational speed of the dynamic antenna array and $\tau$ is the sampling interval in time.
\textcolor{black}{For a rotating dynamic antenna array supporting $N_{\mathrm{r}}=\{1,2,3,...n_{\mathrm{r}}\}$ ring filters, the quantity $D_{\lambda}$ in~(\ref{eq.u_t}) and~(\ref{eq.v_t}) can be simply replaced by $D_{\lambda,\mathrm{rings}}~=~\{D_{\lambda,1},D_{\lambda,2}, D_{\lambda,3},...D_{\lambda,n_{\mathrm{r}}}\}$.}
We note that additional ring filters can be useful to improve image classification by using multiple correlation antenna pair sharing a common centroid for rotation, each tracing out a different ring filter to achieve different $D_{\lambda}$~\cite{9761202}.
%%%%%%%%%%%%%%%%%%%%%%%%%%%%%%%%%%%%%%%%%%%%%%%%%%%%%%%%%%%%%%%%%%%%%%%%%%%%%%%%%%%%%%%%%%%%%%%%%%%%%%%%%%%%%%%%%%%%%%%%%
\begin{figure*}[t!]
	\centering
	\includegraphics[width=\textwidth]{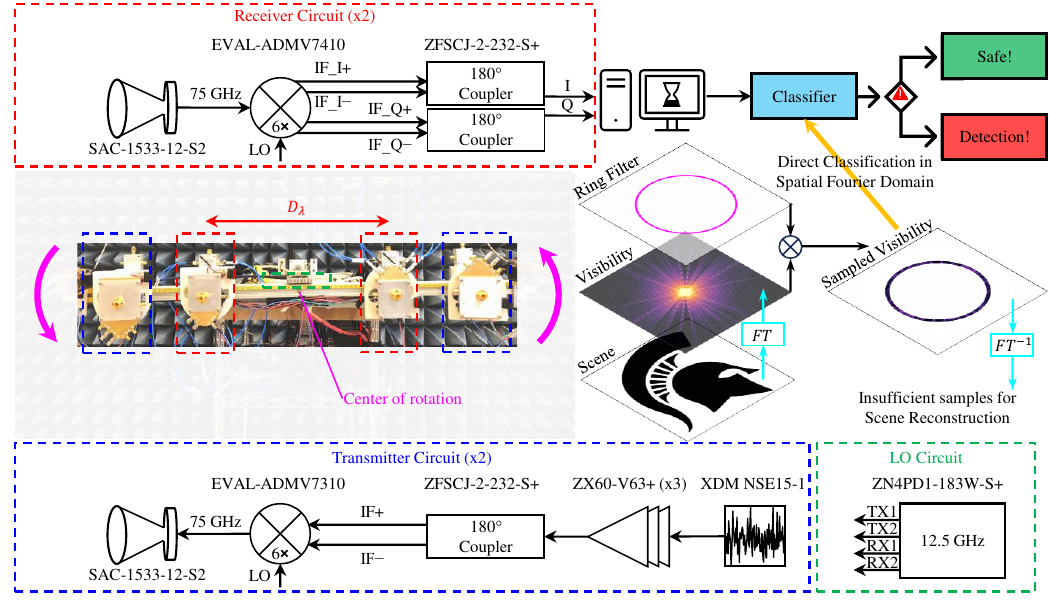}
	\caption{Concept diagram and system architecture of the experimental \SI{75}{GHz} dynamic antenna array. The system includes two transmitters (blue dashed box), two receivers (red dashed box), and local oscillator (LO) (green dashed box). The noise transmitters \textcolor{black}{satisfy} the spatio-temporal incoherence condition which enable Fourier-domain sampling. The two received signals at any given sampled angle are cross-correlated to obtain visibility samples defined by the antenna baseline $D_\lambda$. As the dynamic antenna array rotates, the corresponding Fourier-domain sample also rotates, hence, achieving a ring filter where additional Fourier-domain information can be obtained. The signals are sent to a classifier to determine whether a specific contraband, e.g., a handgun, is concealed by the screened subject. LO: local oscillator. IF: intermediate frequency. TX: transmitter. RX: receiver.}
	\label{fig:SysConArc}
\end{figure*}

\section{Design of Millimeter-Wave Interferometric Dynamic Antenna Array}
Based on the principle of improving unique spatial frequency sampling points through rotational array dynamics, we implemented a two-element interferometric system that operates at \SI{75}{GHz}.
\textcolor{black}{The experimental setup is shown in Fig.~\ref{fig:SysConArc}. The system consists of two noise transmitters and two receivers, all of which are fixed to a rotating arm.}
The two noise transmitters and a two-element interferometric receiver are boxed in dashed-blue and dashed-red, respectively. Both transmitter circuits are identical and both receiver circuits are identical. 
\textcolor{black}{Two transmitters, placed at a wider baseline than the receivers, is necessary to ensure that the radiation incident and reflected from the scene is incoherent in both space and time (further details are in~\cite{9127123}). Additional noise emitters may be used to lower the mutual coherence (obtaining more incoherence) if necessary, which would also lead to an increased SNR, \textcolor{black}{and} enable detection at longer ranges. Note that the downrange performance of the system is not only limited by the SNR but also the spatio-temporal illuminating condition that is dependent on the overlapped radiation patterns from the implemented noise emitters (e.g., spatio-temporal illumination is not maintained if an object is at a location that only receives radiation of a single noise transmitter). Given that a single correlation pair is the smallest sensing unit required to perform Fourier-domain visibility sampling, we implemented the receiving subsystem using two antennas where their centroid is aligned with the dynamic antenna array's center of rotation; subsequent rotational dynamic then generates a ring filter in the spatial Fourier domain as discussed in~(\ref{eq.sf_ring_t}). We note that systems using more than one correlation pair (i.e., multiple ring filters) can benefit from the expanded sampling coverage in the Fourier domain which allows measurements on wider range of spatial frequency responses, and which may improve the accuracy of classification~\cite{9761202}.}

For the transmitter circuits, a noise source of bandwidth between \textcolor{black}{\SI{10}{}\textendash\SI{1600}{MHz}} (RF-Gadgets XDM NSE15-1) is used to satisfy the temporal incoherence aspect (spatial incoherence is accounted for by the placement of the two transmitters outside the baseline of the receivers). The generated noise is amplified through three cascaded baseband amplifiers (Mini-Circuits ZX60-V63+) obtaining a power level of approximately \SI{-4.2}{dBm} that is later fed through a 180$^\circ$ coupler (Mini-Circuits ZFSCJ-2-232-S+). \textcolor{black}{These are used as differential intermediate frequency inputs} of an upconverter with a $6\times$ local oscillator (LO) multiplier (Analog Devices EVAL-ADMV7310) mixing with a LO of \SI{12.5}{GHz}, after which the upconverted noise is radiated through a linearly polarized conical horn antenna (Eravant SAC-1533-12-S2). For the receiver circuits, the linearly polarized conical horn antenna (Eravant SAC-1533-12-S2) is used to capture the spatio-temporally incoherent radiation scattered back from the measured scene where a downconverter with a $6\times$ LO multiplier (Analog Devices EVAL-ADMV7410) is used with a \SI{12.5}{GHz} LO for downcoversion into differential in-phase and quadrature signals where each differential path is combined through a 180$^\circ$ coupler (Mini-Circuits ZFSCJ-2-232-S+), after which the signals are sampled by a digitizer (AlazarTech ATS9416) at a sampling rate of 100M~$\mathrm{Samples}/\mathrm{s}$. The LO circuit boxed by dashed-green in Fig.~\ref{fig:SysConArc} is comprised of a four-way splitter (Mini-Circuits ZN4PD1-183W-S+) where two outputs are for the transmitter circuits, two outputs are for the receiver circuits, and the input local oscillating signal of \SI{12.5}{GHz} is generated using a Keysight E8267D PSG vector signal generator.

Also shown in Fig.~\ref{fig:SysConArc} is the physical implementation of the rotating dynamic antenna array where all transmitting and receiving antennas are co-located on the same rotating hardware to ensure co-polarization. We assume that any depolarization from the scene is negligible. Furthermore, the center of rotation of the dynamic antenna array is coupled to a motor encoder that supports 400 pulses per revolution (PPR), or $360^\circ/400=0.9^\circ$ resolution in terms of the angular rotation. Based on the motor encoder resolution, the digitizer is triggered on the change of angle to measure the scattered radiation from the scene as the dynamic antenna array rotates to synthesize the ring filter. The complete rotational trajectory to synthesize a full ring filter in the spatial frequency domain is $180^\circ$ as the visibility space is Hermitian symmetric because scene intensity is a real quantity. Due to the implemented design, the ring-filtering sampling function of~(\ref{eq.sf_ring_t}) becomes a function of the sampled angles $k$ rather than $t$. This is equivalent to having a uniform sampling interval across the rotating trajectory when, for a sampling interval of $t_{\mathrm{single}}=\tau$ at each measured given angle, the total time required to synthesize a complete ring filter is $t_{\mathrm{ring}} = 200\tau$. Therefore, assuming a constant rotational speed, the speed at which the dynamic antenna rotates will be $\gamma_{\mathrm{ring}}=\tfrac{0.5}{200\tau}$ where a half rotation (i.e., $180^\circ$) accounts for $200\tau$ duration. To ensure that the uniform sampling interval assumption is valid, $\gamma_r\leq\gamma_{\mathrm{ring}}$. When $\gamma_r>\gamma_{\mathrm{ring}}$, certain angles will be skipped making the ring filter incomplete and nonuniform. When all possible spatial frequency samples at all angles are measured, the collection of the cross-correlating outputs can be represented as 
\begin{equation}\label{eq.sf_ring}
	\mathbb{S} = S(u_k,v_k),\;\;\;\forall k =[0, K-1]\mathrm{,}
\end{equation}
where $u_k$ and $v_k$ are the $uv$-samples defined by the two-element array dimension normalized to the wavelength $D_\lambda$ and the $K$ discrete rotational angles $\gamma(k)$ over 180$^\circ$ as $u_k=D_\lambda\sin\gamma(k)$ and $v_k=D_\lambda\cos\gamma(k)$, respectively.

\textcolor{black}{In general, the phase shifts for each receiver chain are important and will require calibration, especially for a system with more than two receiving elements intended for image recovery~\cite{9079644}. Given that a spatial Fourier sample is the cross-correlated output between two receivers, when the number of receivers is greater than two, it is required that the calibrated phase shift among all possible receiver pairs are within tolerance. In our demonstrated setup, the phase difference between the two receivers represents the shift of main beam (from broadside) for the array which was determined negligible.}
%%%%%%%%%%%%%%%%%%%%%%%%%%%%%%%%%%%%%%%%%%%%%%%%%%%%%%%%%%%%%%%%%%%%%%%%%%%%%%%%%%%%%%%%%%%%%%%%%%%%%%%%%%%%%%%%%%%%%%%%%
\section{Imageless Detection of a Real Person Concealing Metallic Gun-shape Object}
The term \textit{subject} refers to the primary background (e.g., a person) carrying an \textit{object} which refers to an item (e.g., a handgun), that can be concealed under clothing. In~\cite{10111012}, the experiment considered a fiber glass mannequin as the subject and a metallic gun-shape object, and presented a heuristically determined threshold approach to differentiate the two measured classes using simple arithmetic statistical features. The results rely on the assumption that the measured scene is static relative to the dynamics of the \SI{75}{GHz} rotational dynamic antenna array. However, this is implausible for a more practical scenario where the screened subject is a real person as small motions due to breathing and/or torso movements can happen at any time throughout the required array dynamics, hence contribute to the degradation of the simple thresh-hold based method which will be discussed below.
%-----------------------------------------------------------------------------------------------------------------------%
\subsection{Experiment Setup}
\begin{figure}[t!]
	\centering
	\subfloat{\includegraphics[width=1.75in]{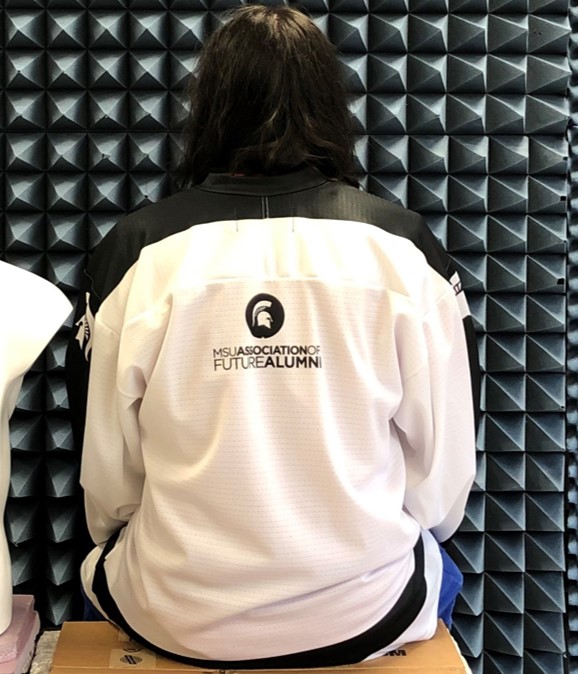}}
	\subfloat{\includegraphics[width=1.75in]{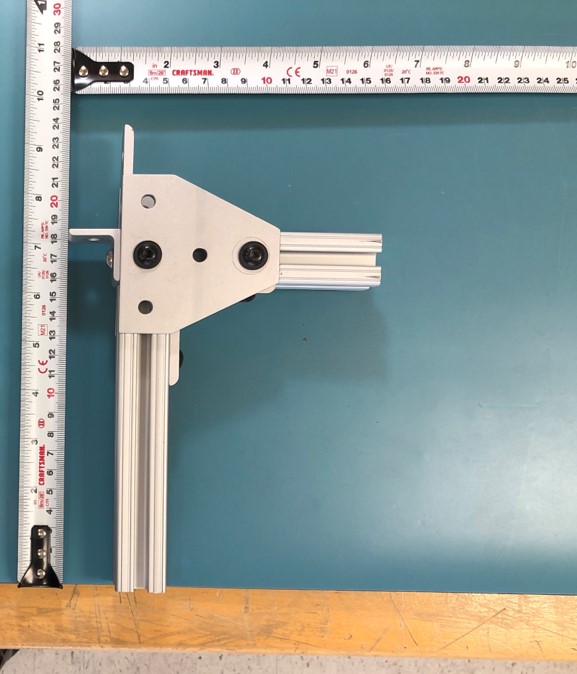}}
	\caption{
		\textbf{Left}: The subject for the imageless contraband detection experiment.
		\textbf{Right}: The object for the imageless contraband detection experiment is a metallic gun-shape with a dimension of \SI{164}{mm}$\times$\SI{235}{mm}.
	}
	\label{fig:sub_obj}
\end{figure}

Experiments were conducted to determine the separability of the Fourier-domain responses when a metallic gun-shape object is concealed under clothing of a real person. The measured subject with and without concealed object as shown in Fig.~\ref{fig:sub_obj} \textcolor{black}{was} approximately \SI{1.83}{m} in front of the dynamic antenna array and backed by walls of radio-frequency absorbers. The concealed object takes the form of a metallic gun-shape that has a dimension of \SI{164}{mm}$\times$\SI{235}{mm}. For the full experiment, the object was randomly placed underneath the clothing with varying orientation between each successive measurement, and the subject is considered to exhibit uncertainties due to the relative orientation to the dynamic antenna array such as breathing and/or torso movements. These considerations represent the varying scattering responses that are measured by the dynamic antenna array.  The experiment consisted of 160 independent measurements equally grouped into two general classes: concealed gun-shape \textcolor{black}{present} versus gun-shape not \textcolor{black}{present}.

On the dynamic antenna array, the two receivers were configured to synthesize a baseline of 77$\lambda$ at \SI{75}{GHz}, and \textcolor{black}{were} rotated over a 180$^\circ$ rotational span at every 0.9$^\circ$ (equivalent to the motor encoder resolution) for all measurements. The two received signals at all 200 angular positions, each with a dwell time of $\tau$=\SI{1}{ms} were then cross-correlated to recover the Fourier information tracing out the ring filtered response $\mathbb{S}$ (i.e., total measuring time of approximately \SI{0.2}{s} per ring). 
%-----------------------------------------------------------------------------------------------------------------------%
\begin{figure*}[t!]
	\centering
	\subfloat{\includegraphics[width=0.2\textwidth]{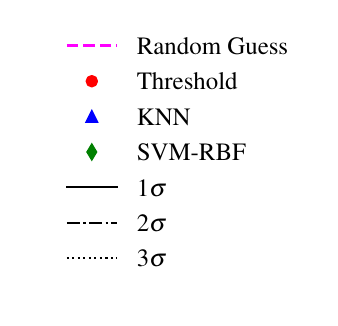}}
	\subfloat{\includegraphics[width=0.2\textwidth]{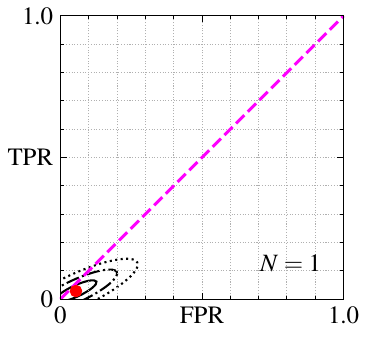}}
	\subfloat{\includegraphics[width=0.2\textwidth]{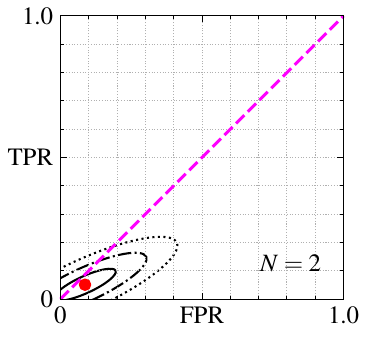}}
	\subfloat{\includegraphics[width=0.2\textwidth]{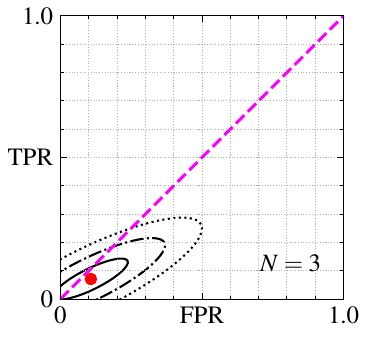}}
	\subfloat{\includegraphics[width=0.2\textwidth]{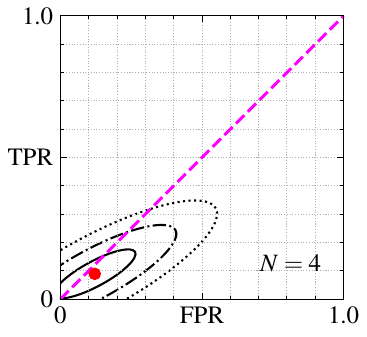}}
	\hfil
	\subfloat{\includegraphics[width=0.2\textwidth]{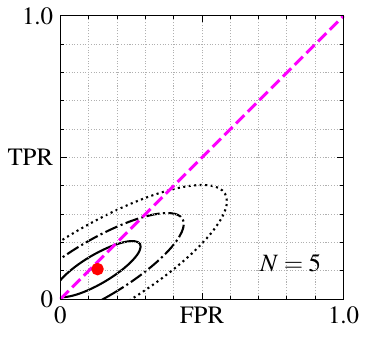}}
	\subfloat{\includegraphics[width=0.2\textwidth]{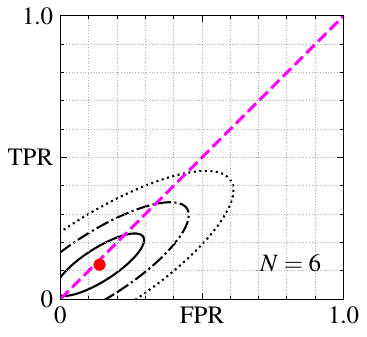}}
	\subfloat{\includegraphics[width=0.2\textwidth]{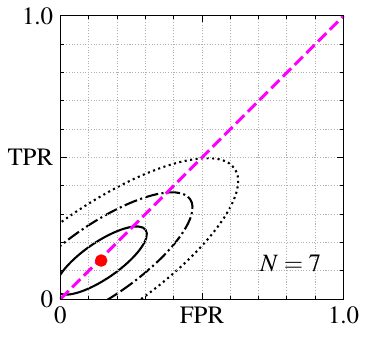}}
	\subfloat{\includegraphics[width=0.2\textwidth]{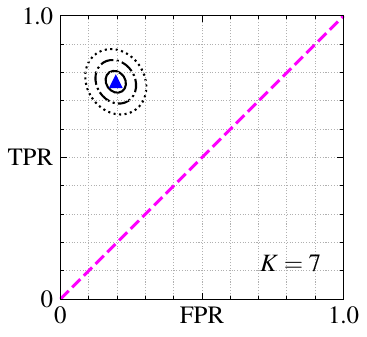}}
	\subfloat{\includegraphics[width=0.2\textwidth]{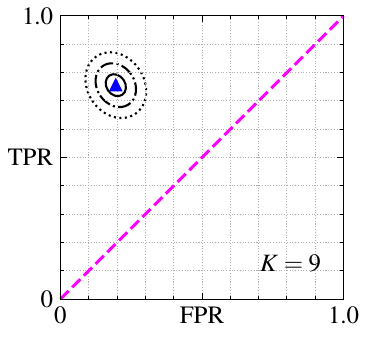}}
	\hfil
	\subfloat{\includegraphics[width=0.2\textwidth]{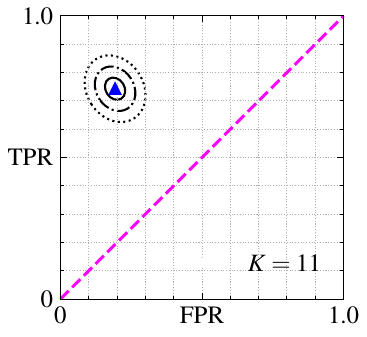}}
	\subfloat{\includegraphics[width=0.2\textwidth]{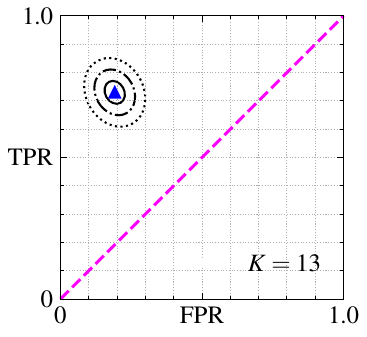}}
	\subfloat{\includegraphics[width=0.2\textwidth]{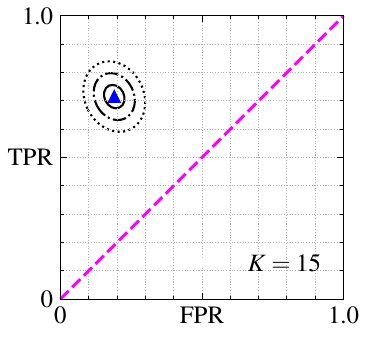}}
	\subfloat{\includegraphics[width=0.2\textwidth]{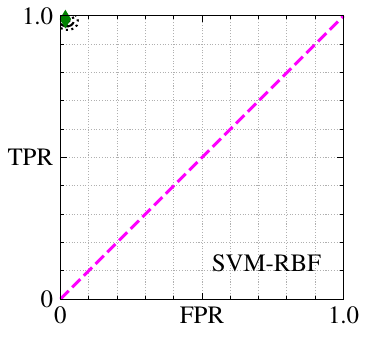}}
	\caption{Receiving operating characteristic (ROC) plots of the 10\,000-iteration Monte Carlo analysis on Experiment~5 using the threshold classifiers (red circle markers), K-nearest neighbor (KNN) classifiers (blue triangle markers), and support vector machine (SVM) using radial basis function (RBF) (green diamond marker). A magenta dashed line is shown to demonstrate the random guess process. The solid black, dash-dot black, and dotted black lines represent the one, two, and three standard deviation ($\sigma$) contours for a particular classifier.}
	\label{fig:exp5_roc_mc}
\end{figure*}
\subsection{Results and Analysis on Data Utility}

A total of 11 heuristically defined features were considered, each a different algorithmic process on the sampled Fourier-domain data $\mathbb{S}$. The 11 features were: mean, median, maximum, standard deviation, variance, difference between maximum and minimum, difference between maximum and mean, difference between maximum and median, difference between mean and minimum, difference between median and minimum, and difference between median and mean. Furthermore, the magnitude of the 11-feature space vector was computed acting as a simple way to reduce the feature space dimension with equal contribution from the 11 features, for each measurement and normalized to all measurements. Subsequently, a 10\,000-iteration Monte Carlo analysis~\cite{10.2307/2280232} was implemented using the simple threshold method based on the heuristically defined features. In each iteration, 70\% of the data were randomly selected and used for training and the remaining 30\% for testing. Furthermore, the number of each class within the training and testing data set were equal. Furthermore, with the assumption that individual measurements are independent, the threshold classifier was applied based on $N$ consecutive measurements for $N=[1,7]$ where $N=1$ and $N=7$ represent classifying a single response and seven consecutive responses, respectively. To evaluate the performance of a classifier, we used the following four metrics~\cite{Altman102, FAWCETT2006861}
\begin{itemize}
	\item True positive rate (TPR) which represents the probability of detection, or successfully identifying an object concealed under the subject's clothing;
	\item False positive rate (FPR) which represents the probability of false detection when only the subject is measured;
	\item Accuracy (ACC) which represents the ratio of all correctly classified data over all data with emphasis on the true positives (i.e., correctly classified subjects with concealed object) and true negatives (i.e., correctly classified subjects without concealed object); and
	\item F1-score (F1) which is a metric similar to ACC but with emphasis on the false negatives (i.e., incorrectly classified subjects with concealed object) and false positives (i.e., incorrectly classified subjects without concealed object);
\end{itemize}
where
\begin{equation}
	\begin{split}
		\mathrm{TPR} = & \frac{\mathrm{TP}}{\mathrm{TP}+\mathrm{FN}}\\
		\mathrm{FPR} = & \frac{\mathrm{FP}}{\mathrm{FP}+\mathrm{TN}}\\
		\mathrm{ACC} = & \frac{\mathrm{TP}+\mathrm{TN}}{\mathrm{TP}+\mathrm{FN}+\mathrm{FP}+\mathrm{TN}}\\
		\mathrm{F1} = & \frac{2\mathrm{TP}}{2\mathrm{TP}+\mathrm{FN}+\mathrm{FP}}\\
	\end{split}
\end{equation}
where TP is the number of true positives (signals correctly identified as contraband), FP is the number of false positives (signals incorrectly identified as contraband), TN is the number of true negatives (signals correctly identified as non-contraband), and FN is the number of false negatives (signals incorrectly identified as non-contraband ). 

\begin{table}[!h]\caption{Averaged classifier metrics based on the reported 10\,000 Monte Carlo simulations shown in Fig.~\ref{fig:exp5_roc_mc}. Bold represents best performing classification scenario based on a single response. THR: Threshold.}\label{tab:exp5_roc_mc}
	\centering
	\begin{tabular}{|c c c c c|}%
		\hline
		\textbf{Scenario} & \textbf{TPR} &  \textbf{FPR} & \textbf{ACC} &  \textbf{F1}\\
		\hline\hline
		\csvreader[late after line=\\\hline]%
		{tables/exp5_roc_mc.csv}{Scenario=\a, TPR=\b, FPR=\c, ACC=\d, F1=\e}%
		{\a & \b & \c & \d & \e}%
	\end{tabular}
\end{table}

\begin{table}[!h]\caption{Standard deviations on the classifier metrics on the reported 10\,000 Monte Carlo simulations shown in Fig.~\ref{fig:exp5_roc_mc}. Bold represents best performing classification scenario based on a single response. THR: Threshold.}\label{tab:exp5_roc_mc2}
	\centering
	\begin{tabular}{|c c c c c|}%
		\hline
		\textbf{Scenario} & $\boldsymbol{\sigma}_\mathrm{TPR}$ &  $\boldsymbol{\sigma}_\mathrm{FPR}$ & $\boldsymbol{\sigma}_\mathrm{ACC}$ &  $\boldsymbol{\sigma}_\mathrm{F1}$\\
		\hline\hline
		\csvreader[late after line=\\\hline]%
		{tables/exp5_roc_mc.csv}{Scenario=\a, sigTPR=\b, sigFPR=\c, sigACC=\d, sigF1=\e}%
		{\a & \b & \c & \d & \e}%
	\end{tabular}
\end{table}

\begin{figure}[t!]
	\centering
	\includegraphics{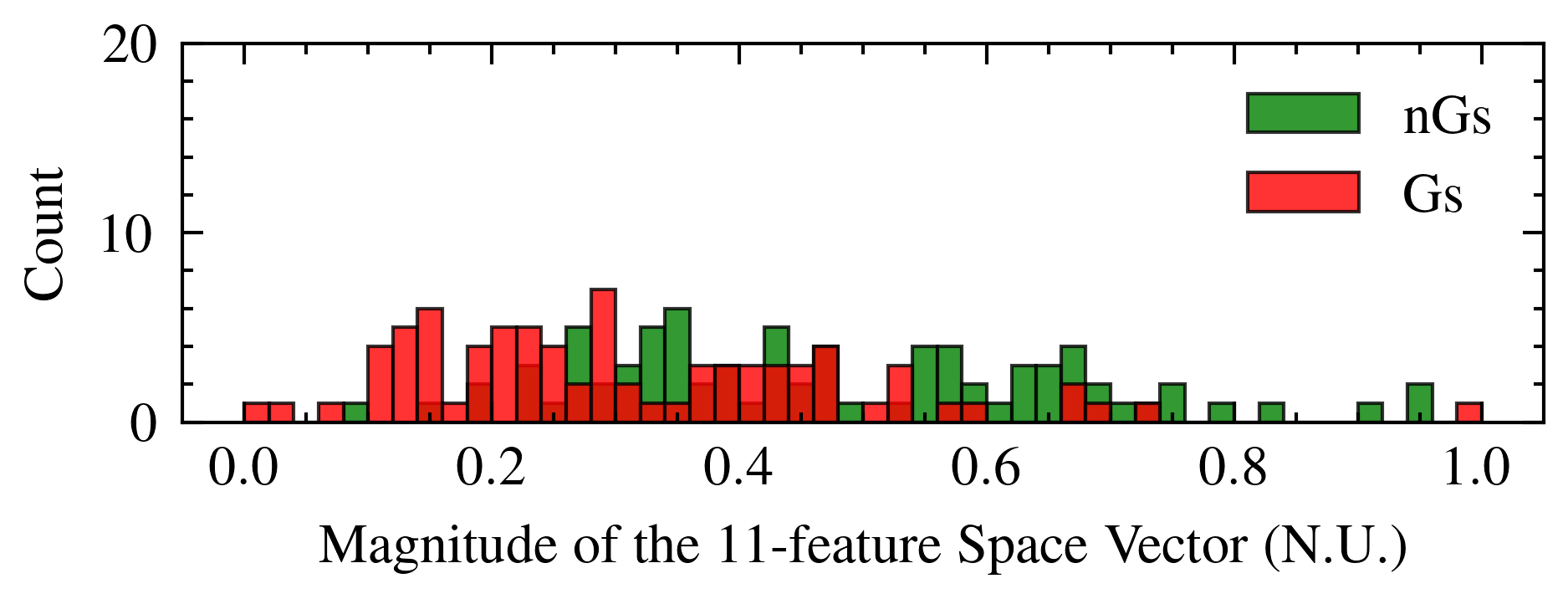}
	\caption{\textcolor{black}{Distribution based on the magnitude of the 11-feature space vector for the two classes Gs (red) and nGs (green) based on measurements of a real person with the back facing towards the dynamic antenna array. Unlike~\cite{10111012}, a plausible threshold value is not apparent for the magnitude of the 11-feature space vector. Gs: Gun-shape. nGs: non-Gun-shape. N.U.: Normalized Unit.}}
	\label{fig:DistAnton}
\end{figure}

The results of the threshold classifiers are shown in Fig.~\ref{fig:exp5_roc_mc} comprising the receiver operating characteristic (ROC) curve~\cite{FAWCETT2006861} that are used to complement the evaluation of a classifier's performance. The ROC curve has two dimensions both ranging from 0 to 1 where the FPR and TPR are associated with the horizontal and vertical dimension, respectively. An ideal classifier will reside exactly at the upper left corner of an ROC curve (FPR~$=0$, TPR~$=1$) suggesting that a classifier's performance can be determined by how close its ROC is to the ideal classifier. On the contrary, the lower right corner of an ROC curve (FPR~$=1$, TPR~$=0$) represent the worst classifier. The diagonal line (magenta-dashed) in Fig.~\ref{fig:exp5_roc_mc} connecting (FPR~$=0$, TPR~$=0$) and (FPR~$=1$, TPR~$=1$) is also important as an ROC value residing on this line is equivalent to a classifier that is based on a random guess process, hence, a good classifier should always be above and away from this line. As observed, the threshold classifier failed to differentiate between the two measured classes regardless of the consideration of $N$ consecutive measurements and that the outcome is similar to a random guess process as shown in Fig.~\ref{fig:exp5_roc_mc}, and summarized in Table~\ref{tab:exp5_roc_mc} and Table~\ref{tab:exp5_roc_mc2}. We note the degradation of the threshold classifier used in~\cite{10111012} can be related with motion of the person during the measurements. While~\cite{10111012} \textcolor{black}{considered} varying position and orientation of the measured subject and object, the scattering profile remains constant for any given measurement. This is no longer valid when the measured subject is a real person \textcolor{black}{because} a single measurement can capture various scattering profiles due to the person's movements. \textcolor{black}{As shown in Fig.~\ref{fig:DistAnton}, unlike the reported results considering a static mannequin~\cite{10111012}, a plausible threshold value is not apparent, which explains the poor performance of the threshold approach.}

In addition, one inevitable challenge with the implemented threshold classifier used in~\cite{10111012} is the contribution of outliers in the training data set causing the exact value of the threshold to be significantly shifted, therefore degrading the classification. Another potential issue is when the two considered classes are not linearly separable in the training feature space. This could be due to an extremely complex decision boundary or the boundary might not exist. In addition, while classifying $N$ consecutive responses provides opportunity to improve the overall performance, this approach only works when a classifier's performance on a single response is robust; and classifying on multiple consecutive responses can lead to longer screening time which can be undesirable. We note that a shorter sampling interval $\tau$ \textcolor{black}{enabled} by sufficient receiving signal-to-noise ratio (SNR), can allow the dynamic antenna array to rotate at a faster speed such that the total measurement time is shorter than \textcolor{black}{periodicity} of common motions of a person (e.g., breathing and shivering). However, this does not guarantee that motion of a person will not be \textcolor{black}{present} during the duration $\tau$. Therefore, we consider two new classifiers as potential alternatives to address the above concerns and to complement our investigation of the privacy preserving contraband detection technique. The first classifier considered is the $K$-nearest neighbor (KNN) classifier~\cite{1053964, pattern}. For an incoming unknown data point, the KNN classifier operates by considering $K$-nearest data points in the training feature space that are local to the unknown point where the classification outcome is based on the majority class type among the $K$-nearest neighboring points in the training data set. We note that while KNN can be more robust to potential outliers in the training data set as it benefits from the clustering effect of the data points rather than drawing specific decision boundary, it has an inherent drawback of higher computational expense due to the process of evaluating the nearest neighboring points as well as additional computational cost with increasing $K$ values. The second considered classifier is the support vector machine (SVM) using radial basis function (RBF)~\cite{Cortes1995, 650102, andrew_2000}. SVM is a technique that seeks to find the best hyperplane (i.e., decision boundary) that maximizes the distance between the training points of the classes. When the two considered classes are not linearly separable in the original feature space, a kernel function is typically applied to transform the original feature space into a higher-dimensional feature space where the points between the two classes become linearly separable. One such kernel function is the RBF that is also commonly known as the Gaussian radial basis kernel. During the classifier training stage, the SVM-RBF determines the proper regularization parameters such that the width of the kernel function (i.e., RBF) is neither too wide nor too narrow which controls how much influence the training points have on the overall classifier. Once an RBF is determined, the SVM algorithm identifies the best hyperplane in the higher-dimensional feature space. During the classification stage, an incoming unknown data is transformed to the higher-dimensional feature space by the trained RBF and classify against the trained hyperplane.

The same Monte Carlo simulation configuration (i.e., number of iterations and training-testing data split) was applied to both the KNN and SVM-RBF classifiers. For the KNN classifier, five values of considered neighbors $K=[7,9,11,13,15]$ were used. For the SVM-RBF classifier, the regularization parameters controlling the RBF are determined per iteration using a logarithmic grid search approach to maximize the accuracy of the classifier based on the training data set. The corresponding results are shown in Fig.~\ref{fig:exp5_roc_mc} and summarized in Table~\ref{tab:exp5_roc_mc} and Table~\ref{tab:exp5_roc_mc2}. Unlike the threshold based method, both the KNN classifier (for all considered $K$) and the SVM-RBF classifiers achieved comparably better performance. The SVM-RBF method is the best performing classifier with an ACC~$=0.986$ and F1~$=0.986$ considering its ROC of FPR~$=0.017$ and TPR~$=0.989$. Furthermore, it is also noted that the SVM-RBF classifier exhibit the smallest standard deviation across the Monte Carlo simulation followed by the KNN classifier then the randomly guessing threshold classifier.
%-----------------------------------------------------------------------------------------------------------------------%
\begin{table}[!t]\caption{Averaged elapsed time for individual processes of the demonstrated imageless contraband detection technique. THR: Threshold.}\label{tab:processTimes}
	\centering
	\begin{tabular}{|c c|}%
		\hline
		\textbf{Process} & \textbf{Time (ms)} \\
		\hline\hline
		\csvreader[late after line=\\\hline]%
		{tables/processTimes.csv}{process=\a, time=\b}%
		{\a & \b}%
	\end{tabular}
\end{table}

\begin{table*}[!t]\caption{Comparison of techniques intended for contraband detection application. TX: Transmitter. RX: Receiver. ACC: Accuracy.}\centering{
		\begin{tabular}{|c|c|c|c|c|c|c|}
			\hline
			%			\textbf{Work} & \textbf{\begin{tabular}[c]{@{}c@{}}Frequency\\ (GHz)\end{tabular}} & \textbf{\begin{tabular}[c]{@{}c@{}}Number of\\ Antennas\end{tabular}} & \textbf{\begin{tabular}[c]{@{}c@{}}Image\\ Formation\end{tabular}} & \textbf{\begin{tabular}[c]{@{}c@{}}Detection\\ Demonstration\end{tabular}} & \textbf{\begin{tabular}[c]{@{}c@{}}Number of\\ Detected Classes\end{tabular}} & \textbf{\begin{tabular}[c]{@{}c@{}}Processing\\ Time\end{tabular}} \\ \hline\hline
			\shortstack{\textbf{Referencing}\\\textbf{Work}}
			& \shortstack{\textbf{Frequency}\\\textbf{(GHz)}} 
			& \shortstack{\textbf{Number~of}\\\textbf{Antennas}}
			& \shortstack{\textbf{Image}\\\textbf{Formation}}
			& \shortstack{\textbf{Detection}\\\textbf{Demonstration}}
			& \shortstack{\textbf{Number~of}\\\textbf{Detected~Classes}}
			& \shortstack{\textbf{Processing}\\\textbf{Time}}\\ \hline\hline
			\cite{6651696}       & 20\textendash26     & 40 & Yes & No  & n.a. & \SI{15}{s} \\ \hline
			\cite{8466794}       & 20\textendash30     & 256 & Yes & No (Visual) & n.a. & \SI{1.5}{s} \\ \hline
			\cite{6080744}       & 72\textendash80     & \begin{tabular}[c]{@{}c@{}}1472\\ (736 TX + 736 RX)\end{tabular}      & Yes & No (Visual) & n.a. & \SI{157}{ms} \\ \hline
			\cite{9564597}       & 32                  & 220                                                                   & Yes & No (Visual) & n.a. & \SI{2.07}{s} \\ \hline
			\cite{9521903}       & 12                  & \begin{tabular}[c]{@{}c@{}}2\\ (1 TX + 1 RX)\end{tabular}             & Yes & \begin{tabular}[c]{@{}c@{}}Yes\\ (Best ACC: 0.930)\end{tabular} & Four & \SI{3.8}{ms} \\ \hline
			This Work     & 75                         & \begin{tabular}[c]{@{}c@{}}4\\ (2 TX + 2 RX)\end{tabular}             & No & \begin{tabular}[c]{@{}c@{}}Yes\\ (Best ACC: 0.986)\end{tabular} & Two & \SI{211}{ms} \\ \hline
		\end{tabular}
	}\label{tab:comparison}
\end{table*}

\begin{figure}[h!]
	\centering
	\subfloat{\includegraphics[width=0.5\linewidth]{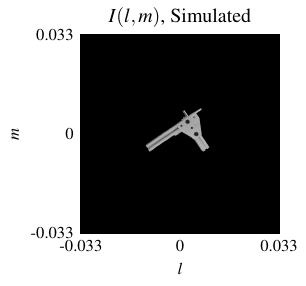}}
	\subfloat{\includegraphics[width=0.475\linewidth]{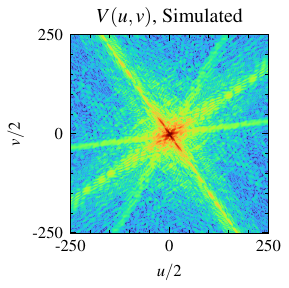}}
	\hfil
	\subfloat{\includegraphics[width=0.5\linewidth]{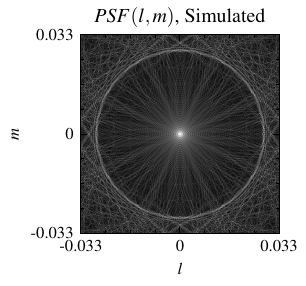}}
	\subfloat{\includegraphics[width=0.475\linewidth]{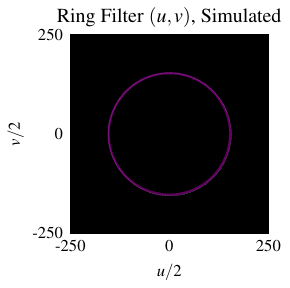}}
	\hfil
	\subfloat{\includegraphics[width=0.5\linewidth]{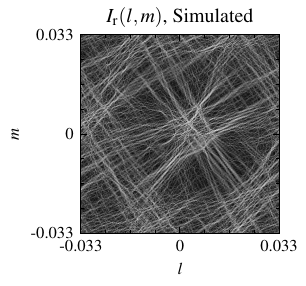}}
	\subfloat{\includegraphics[width=0.5\linewidth]{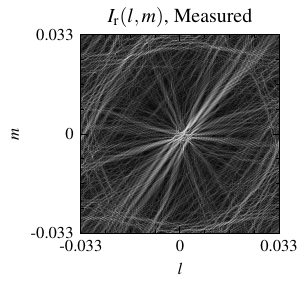}}
	\caption{
		\textbf{Top Row}: Simulated scene intensity using the metallic gun-shape object in the imageless contraband detection measurements (left), and the simulated visibility (right).
		\textbf{Center Row}: Simulated point spread function (PSF) of the ring filter (i.e., sampling function) based on the receivers configuration of the rotational dynamic antenna array used in the imageless contraband detection measurement. Note that a semi-transparent magenta annotation is included in the ring filter plot to illustrate spatial Fourier regions that the ring filter samples.
		\textbf{Bottom Row}: The simulated unrecoverable scene intensity based on the simulated ring filtered visibility (i.e., product between the simulated ring filter and simulated visibility) of the metallic gun-shape object (left). The measured unrecoverable scene intensity of the metallic gun-shape object using the same systems setup as in the imageless contraband detection measurements where the object is horizontally and vertically aligned to the array's center of rotation (right).
	}
	\label{fig:sim_vs_meas}
\end{figure}

\begin{figure}[h!]
	\centering
	\subfloat{\includegraphics[width=0.5\linewidth]{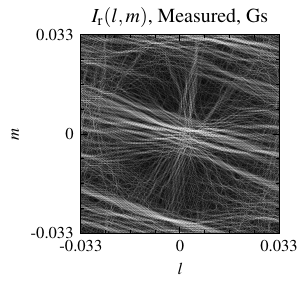}}
	\subfloat{\includegraphics[width=0.5\linewidth]{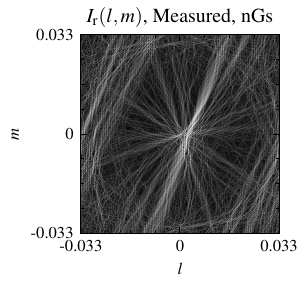}}
	\hfil
	\subfloat{\includegraphics[width=0.5\linewidth]{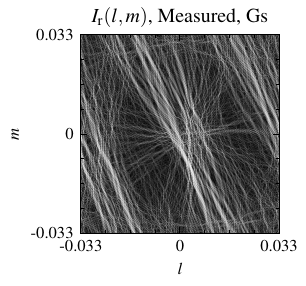}}
	\subfloat{\includegraphics[width=0.5\linewidth]{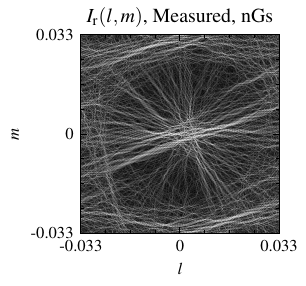}}
	\hfil
	\subfloat{\includegraphics[width=0.5\linewidth]{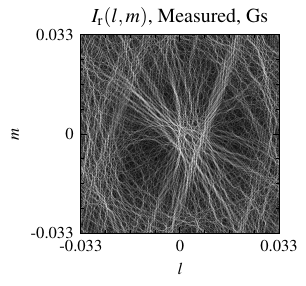}}
	\subfloat{\includegraphics[width=0.5\linewidth]{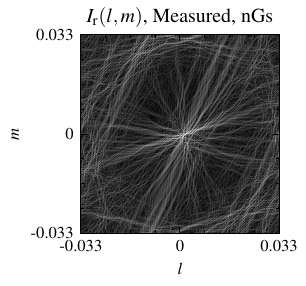}}
	\caption{Examples demonstrating unrecoverable image reconstruction based on the ring-filtered visibilities for the imageless concealed contraband detection of a real person with (left column) and without the metallic gun-shape (right column). Gs: Gun-shape (with real person). nGs: No gun-shape (real person only).}
	\label{fig:meas_unrecoverable}
\end{figure}

\begin{figure}[h!]
	\centering
	\subfloat{\includegraphics[width=0.5\linewidth]{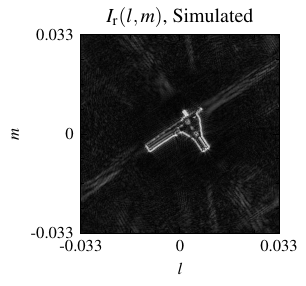}}
	\subfloat{\includegraphics[width=0.475\linewidth]{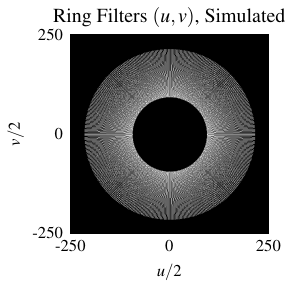}}
	\caption{Simulated examples showing that the rotational dynamic antenna array can recover useful spatial information when sufficient spatial Fourier information are measured such as using the technique demonstrated in~\cite{9827908}.
	}
	\label{fig:recoverable}
\end{figure}

\subsection{Processing Cost Analysis}
The ability for real-time operation is one important consideration for a screening system to be used for real world application. The imageless detection system processing can be generalized to four separate processes (in sequence): data acquisition, visibility generation, feature extraction, and classification (i.e., inference). Averaged values of the computational time of each process are shown in Table~\ref{tab:processTimes}; the processor of the host machine was an Intel\textregistered~Core\texttrademark~i9-9820X. The data acquisition time was \SI{200}{ms} and accounts for the majority of the duration from measurement to inference. We note that the resolution of the angle rotation, the dwell time per angle, and the integration time, among other factors, contribute to the data acquisition time required to complete a single screening measurement. The visibility generation time was \SI{8.53}{ms} for a single screening measurement with 200 angles. The feature extraction time to obtain the 11 features was \SI{0.03}{ms}. The time to infer an incoming unknown sample ranged from \textcolor{black}{\SI{0.22}{}\textendash\SI{2.01}{ms}} depending on the classifier. The threshold method (for a single event) required the least amount of time followed by SVM-RBF, and followed by KNN in increasing $K$ values which is expected for KNN classifier where the choice of $K$ affects the number of neighboring data that will be compared. Based on Table~\ref{tab:processTimes}, the longest duration from measurements to inference was approximately \SI{211}{ms} considering the KNN (K=15) scenario. As discussed in~\cite{6080744}, a measurement time of a few hundred milliseconds is sufficient for screening people that are standing or sitting still. Note that the above assumes a serial configuration for the four processes and that the training and evaluation for the classifier is done offline. Furthermore, the demonstrated technique can continue to operate for screening without reset due to the benefit of the rotational array dynamics. In Table~\ref{tab:comparison}, we compared the demonstrated imageless contraband detection technique to image-based techniques that are also intended for contraband detection application~\cite{6651696, 8466794, 6080744}. Nevertheless, it is worth noting that the presented imageless technique accounts for the full screening process from measurements to inference in under \SI{211}{ms} which is feasible for real-time applications, considering that some techniques consume more time simply for image formation~\cite{6651696,8466794}. Furthermore, current airport security screening systems can take up to \SI{1.5}{s} to complete scanning and up to \SI{6}{s} to complete both scanning and detection~\cite{leidos,9318765}. Finally, it is evident that the presented imageless contraband detection approach enables significant hardware reduction (i.e., number of antennas).

%-----------------------------------------------------------------------------------------------------------------------%

\subsection{Privacy Preservation via Unrecoverable Image Reconstruction}
%From the above analyses, we demonstrated the utility of the ring-filtered responses. 
In this section, we demonstrate the inherent privacy preservation attribute of the demonstrated technique. As shown in the top row of Fig.~\ref{fig:sim_vs_meas}, we simulated a reference scene and its visibility using the metallic gun-shape object described in the above experiment. In the center row of Fig.~\ref{fig:sim_vs_meas}, the simulated point spread function (PSF) \textcolor{black}{and ring filter are shown where the former is based on its two-dimensional Fourier transform pair (i.e., the sampling function, ring filter)~\cite{Thompson2017}}. A semi-transparent magenta annotation is included in the ring filter plot to illustrate spatial Fourier regions that the ring filter samples. In the bottom row of Fig.~\ref{fig:sim_vs_meas}, the simulated scene intensity reconstruction is shown which is processed based on the interferometric imaging technique as described in (\ref{eq.ir}). As observed, the reconstructed scene intensity can be considered to be unrecoverable since no perceptible spatial information are represented in the reconstructed image. To demonstrate the fact that the recovered "image" does not represent useful data from which biometrics may be obtained when sensing a person, we computed the structural similarity index measured (SSIM) between the simulated reference and the reconstruction. The SSIM is an image-specific quality metric that considers structural information, contrast, and luminance between a reference and the recovered image; and it has a range of $[-1,1]$ where 1 represents identical images, and 0 and -1 indicate no similarity and perfect anti-correlation, respectively~\cite{1284395}. The reconstructed image yielded a SSIM with a value of 0.070, meaning that there is essentially no correlation between the information in the reconstructed image and that in the original image. This demonstrates that the sensing technique does not provide sufficient information from which biometric data may be obtained when sensing a person.

In the bottom right of Fig.~\ref{fig:sim_vs_meas}, we present a measured unrecoverable scene intensity of the metallic gun-shape object using the same systems setup as in the imageless contraband detection measurements where the object is horizontally and vertically aligned to the array's center of rotation with similar orientation as the simulated reference scene (top left of Fig.~\ref{fig:sim_vs_meas}). Similar to the simulated reconstruction, the measured reconstruction yielded no useful spatial information demonstrating the capability for privacy preservation. 
In addition, we include three unrecoverable reconstruction examples for each of the two classes from the imageless classification measurements on a person in Fig.~\ref{fig:meas_unrecoverable}. Evidently, the recovered measured scenes exhibit no spatial information of the screened person nor the gun-shape object. 

To further illustrate the privacy-preserving aspect, we conducted a simulation using a similar rotational Fourier-domain system, however one that includes far more samples and that is intended to reconstruct images; we demonstrated such a system in~\cite{9827908}. Fig.~\ref{fig:recoverable} shows the reconstructed gun-shaped object shown on the left-hand side of the figure obtained from a far denser sampling function shown on the right-hand side of the figure. It is evident that the proposed approach, using far fewer spatial frequency samples, does not collect sufficient information to form images, and is thus inherently privacy-preserving.
%%%%%%%%%%%%%%%%%%%%%%%%%%%%%%%%%%%%%%%%%%%%%%%%%%%%%%%%%%%%%%%%%%%%%%%%%%%%%%%%%%%%%%%%%%%%%%%%%%%%%%%%%%%%%%%%%%%%%%%%%
\section{Conclusion}
We experimentally demonstrated a technique to detect concealed contraband while achieving privacy preservation of the screened subject by preventing image reconstruction through measuring a substantially reduced set of information enabled by a rotating dynamic antenna array operating at \SI{75}{GHz}. The experimental setup uses commercially available components and classifiers operating on simple arithmetic statistical features extracted from the direct measured Fourier-domain responses, and achieves a classification accuracy and F1-score both well above 0.900 when considering a real person as the screened subject with and without concealing a gun-shaped object beneath clothing. While this work considered metal gun-sized contraband objects, detection of dielectric and smaller objects is expected to be feasible with sufficient SNR. The results validate prior predictions on the application of real-time imageless contraband detection as well as demonstrate the feasibility and robustness of the presented approach considering the different subject and/or object combinations evaluated. While previous microwave and millimeter-wave contraband detection systems have relied on image formation, this work demonstrates the feasibility of detecting objects without image formation. Furthermore, the approach uses less hardware and has lower computational complexity, providing an efficient means for objects detection. Both aspects may prove useful for advancing the presented concept toward real world scenarios\textcolor{black}{, accounting for a wider range of screening scenarios, when paired} with machine-learning-based classifiers, more sophisticated feature extraction techniques as well as other complementing privacy enhancing techniques.
%%%%%%%%%%%%%%%%%%%%%%%%%%%%%%%%%%%%%%%%%%%%%%%%%%%%%%%%%%%%%%%%%%%%%%%%%%%%%%%%%%%%%%%%%%%%%%%%%%%%%%%%%%%%%%%%%%%%%%%%%
\bibliographystyle{ieeetran}
\bibliography{IEEEabrv,Daniel_Bib}
%%%%%%%%%%%%%%%%%%%%%%%%%%%%%%%%%%%%%%%%%%%%%%%%%%%%%%%%%%%%%%%%%%%%%%%%%%%%%%%%%%%%%%%%%%%%%%%%%%%%%%%%%%%%%%%%%%%%%%%%%
%If you do not have or do not want to include a photo, you can use IEEEbiographynophoto as shown below:
\begin{IEEEbiography}[{\includegraphics[width=1in,height=1.25in,clip,keepaspectratio]{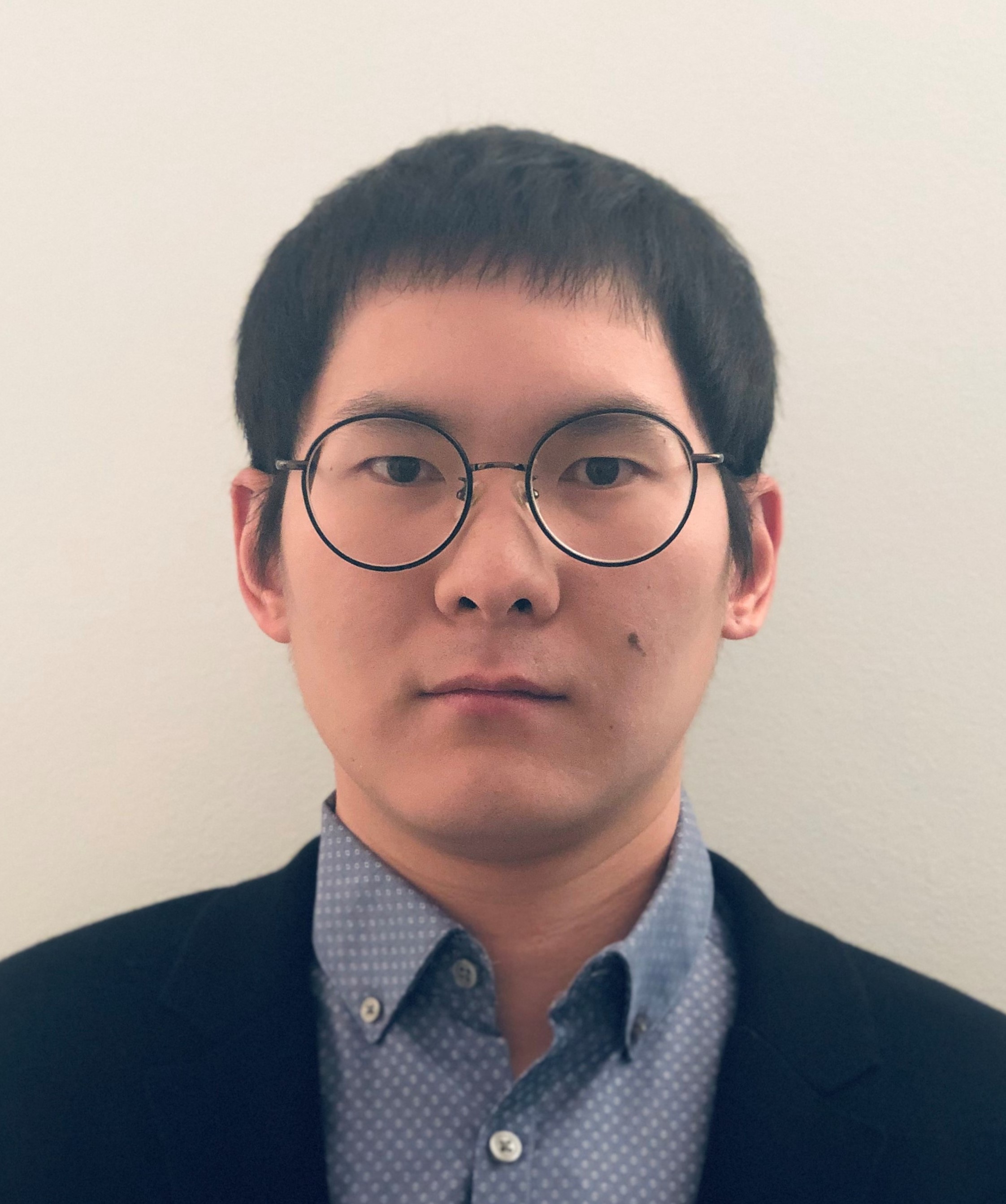}}]{Daniel Chen}
	(Graduate Student Member, IEEE) received the B.S. and M.S. degree in electrical engineering from Michigan State University, East Lansing, MI, USA, in 2015 and 2019, respectively. He is currently pursuing the Ph.D. degree in electrical and computer engineering at Michigan State University, East Lansing, MI, USA. His current research interests include wireless microwave and millimeter-wave systems, antenna arrays, radars and signal processing.
\end{IEEEbiography}

\begin{IEEEbiography}[{\includegraphics[width=1in,height=1.25in,clip,keepaspectratio]{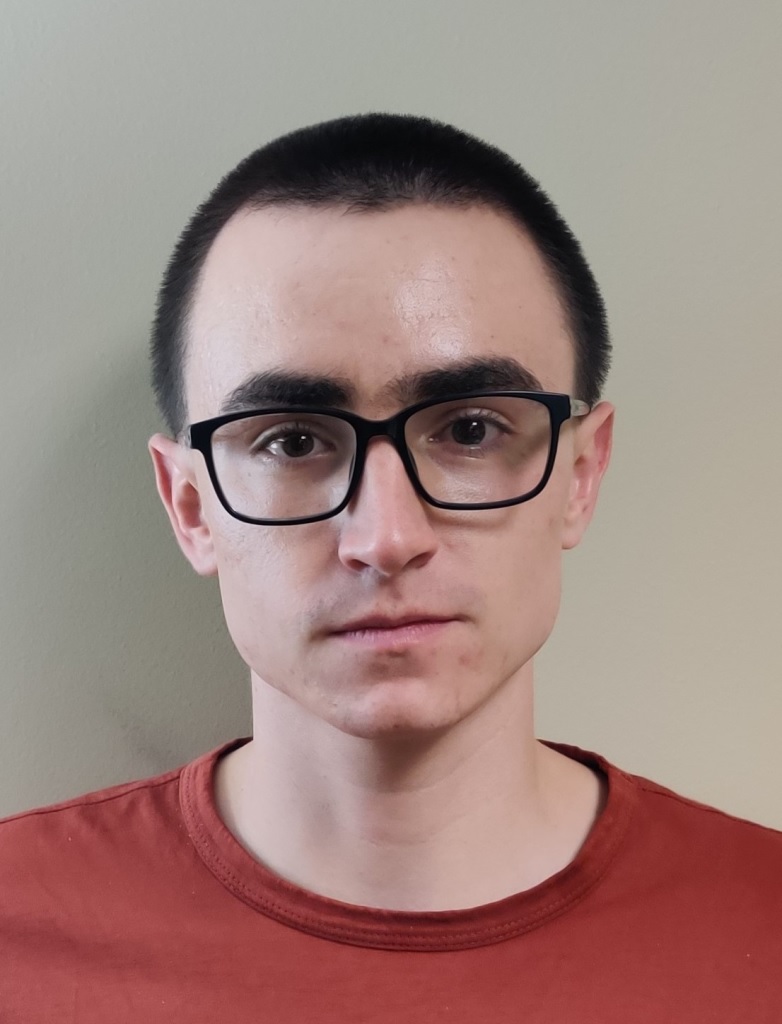}}]{Anton Schlegel}
 (Graduate Student Member, IEEE) received the B.S. degree in computer engineering from Michigan State University, East Lansing, MI, USA, where he is currently pursuing the Ph.D. degree in electrical engineering.,His research interests include wireless microwave and millimeter-wave systems, distributed antenna arrays, and signal processing.
\end{IEEEbiography}

\begin{IEEEbiography}[{\includegraphics[width=1in,height=1.25in,clip,keepaspectratio]{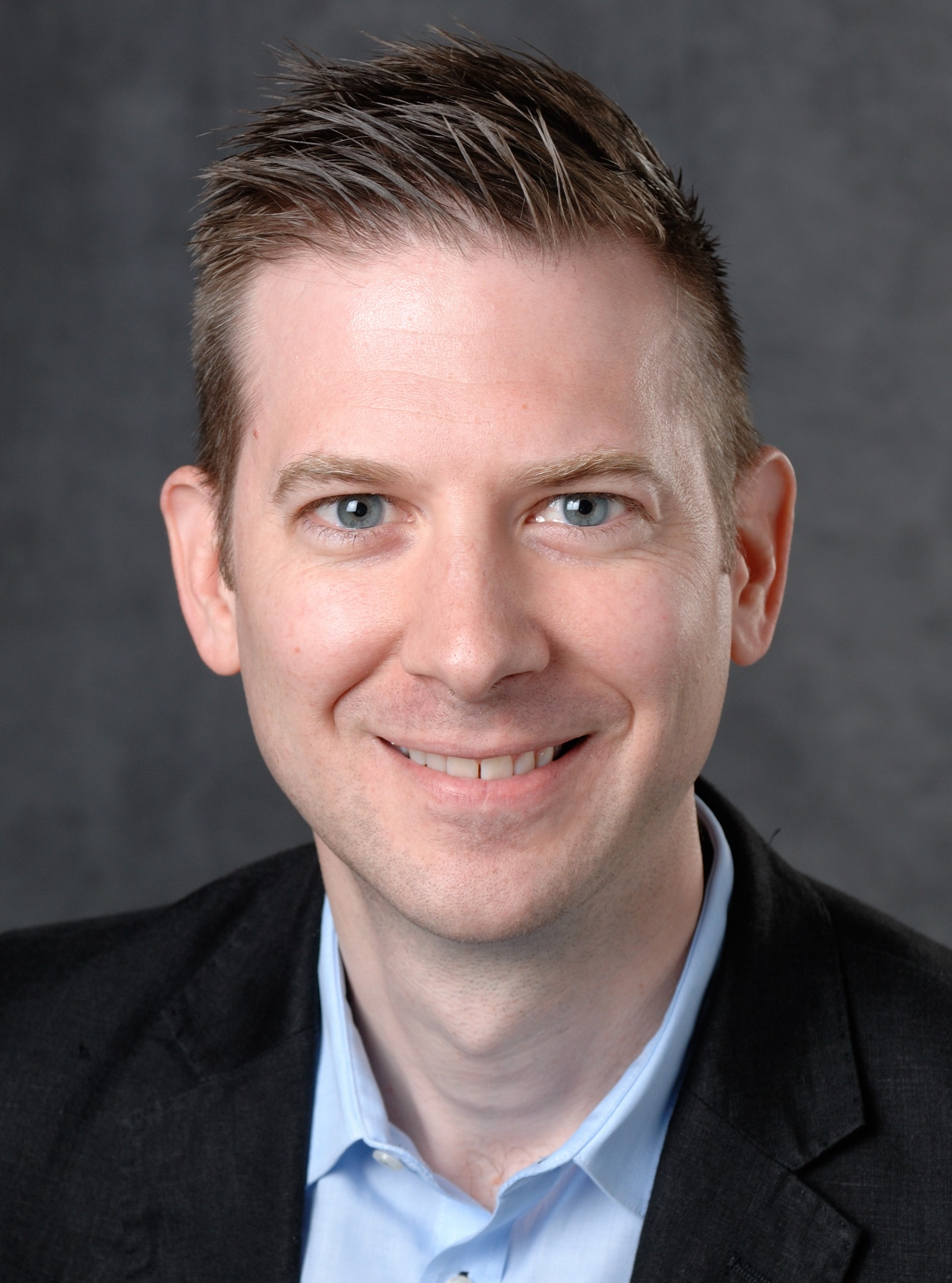}}]{Jeffrey A. Nanzer} (S'02--M'08--SM'14) received the B.S. degrees in electrical engineering and in computer engineering from Michigan State University, East Lansing, MI, USA, in 2003, and the M.S. and Ph.D. degrees in electrical engineering from The University of Texas at Austin, Austin, TX, USA, in 2005 and 2008, respectively.
	
	From 2008 to 2009 he was with the University of Texas Applied Research Laboratories in Austin, Texas as a Post-Doctoral Fellow designing electrically small HF antennas and communications systems. From 2009 to 2016 he was with the Johns Hopkins University Applied Physics Laboratory where he created and led the Advanced Microwave and Millimeter-Wave Technology Section. In 2016 he joined the Department of Electrical and Computer Engineering at Michigan State University where he held the Dennis P. Nyquist Assistant Professorship from 2016 through 2021. He is currently an Associate Professor. He directs the Electromagnetics Laboratory, which consists of the Antenna Laboratory, the Radar Laboratory, and the Wireless Laboratory. He has published more than 200 refereed journal and conference papers, two book chapters, and the book  Microwave and Millimeter-Wave Remote Sensing for Security Applications (Artech House, 2012). His research interests are in the areas of distributed phased arrays, dynamic antenna arrays, millimeter-wave imaging, remote sensing, millimeter-wave photonics, and electromagnetics.
	
	Dr. Nanzer is a Distinguished Microwave Lecturer for the IEEE Microwave Theory and Techniques Society (Tatsuo Itoh Class of 2022-2024). He was a Guest Editor of the Special Issue on Special Issue on Radar and Microwave Sensor Systems in the IEEE Microwave and Wireless Components Letters in 2022, and from 2017-2023 was an Associate Editor of the IEEE Transactions on Antennas and Propagation. He is a member of the IEEE Antennas and Propagation Society Education Committee and the USNC/URSI Commission B, was a founding member and the First Treasurer of the IEEE APS/MTT-S Central Texas Chapter, served as the Vice Chair for the IEEE Antenna Standards Committee from 2013 to 2015, and served as the Chair of the Microwave Systems Technical Committee (MTT-16), IEEE Microwave Theory and Techniques Society from 2016 to 2018. He was a recipient of the Withrow Distinguished Junior Scholar Award in 2024, the Google Research Scholar Award in 2022 and 2023, the IEEE MTT-S Outstanding Young Engineer Award in 2019, the DARPA Director’s Fellowship in 2019, the National Science Foundation (NSF) CAREER Award in 2018, the DARPA Young Faculty Award in 2017, and the JHU/APL Outstanding Professional Book Award in 2012.  
\end{IEEEbiography}

\EOD
\end{document}